\documentclass[
    aip,
    pop,
    reprint,
    amsmath,
    amssymb,
    superscriptaddress
]{revtex4-2}

\usepackage{graphicx}   
\usepackage{dcolumn}    
\usepackage{bm}         
\usepackage{amsmath}
\usepackage{amssymb}
\usepackage{siunitx}    
\usepackage{hyperref}  
\usepackage{color}
\begin{document}

\title{Electric field-driven Rayleigh Taylor-like instability in binary complex plasma}

\author{Priya Deshwal}
\email {priyadeshwal25@gmail.com} 

\author{Hitendra K. Malik}
\affiliation{Physics Department, IIT Delhi, Hauz Khas, New Delhi - 110016, India}

\begin{abstract}
This study uses two-dimensional molecular dynamics simulations to explore Rayleigh-Taylor-like instability in a strongly coupled binary complex plasma, when heavier dust particles are positioned above the lighter ones, having the same charge-to-mass ratio, in a planar configuration. Langevin dynamics simulations are used to study the evolution of perturbations at the interface between these two distinct species of charged dust particles, subject to an externally applied electric field and an equilibrium charge density gradient. We have performed analytical calculations to determine the growth rate of instability under both strongly coupled and weakly coupled dusty plasma regimes. These theoretical predictions are subsequently validated through molecular dynamics simulations, enabling the reproduction of the instability in a strongly coupled binary complex plasma. The instability in these cases ultimately causes mixing among the charged species. The study comprehensively analyzes the growth rate as a function of various system parameters, and it offers deeper insight into the underlying physical mechanisms.

\end{abstract}

\maketitle

 \section{\it Introduction}
 
Dusty plasma is an ionized gas containing solid particles of various sizes, either induced externally or produced inside the plasma, which get charged by attracting either electrons or ions from the plasma. Generally, electrons are more mobile species and attach to the dust particles, making them negatively charged. The high charge on dust particles increases the intergrain interaction energy compared to the average thermal kinetic energy. The ratio of these two types of energies is the Coulomb coupling parameter, $\Gamma = Q^2/4\pi\epsilon_0 a k_B T $, where Q, a, and T represent the charge, intergrain distance, and temperature of dust species, respectively. Hence, achieving the strong coupling condition involves (i) increasing the charge Q, (ii) minimizing the distance between particles a, and (iii) lowering the temperature T. Depending on the value of $\Gamma$, this medium can be categorized as a gas, liquid, and solid. The dust grains interact with each other via a screened Coulomb potential $v(r) = (Q/4\pi \epsilon_0 r) exp(-r/\lambda_d)$ \cite{konopka2000measurement,lampe2000interactions}  instead of the Coulomb potential, where $\lambda_d$ is the Debye length. Thus, it is a complex medium that can be understood with minimal reference to background particles. In the field of basic plasma physics, dusty plasma is a versatile medium to study waves \cite{barkan1996experiments,rao1990dust,shukla1992dust} and instabilities \cite{ma1994self,rosenberg1996ion} and Coulomb crystallization \cite{melzer1994experimental}; also the formation of plasma crystal \cite{thomas1994plasma}, which has been shown to exhibit phase transitions \cite{pieper1996experimental,morfill1996plasma}.

  The charge-to-mass ratio of dust species makes the time scale associated with dust species slow compared to the time scales of electrons and ions; as a result, electrons and ions can be assumed to be inertia-less on the dust time scale. Thus, their dynamics can be easily visualized and captured even by normally charged coupled devices. Thus, dusty plasma is a test-bed medium to study linear and nonlinear waves \cite{rao1990dust, shukla1992dust}, instabilities \cite{merlino1998laboratory, d1993rayleigh, d1990kelvin}, compressional waves \cite{pieper1996dispersion}, and collective properties \cite{kaw1998low, kaw2001collective, ivlev2000anisotropic}. 
Dusty plasma can be easily prepared in the strongly coupled state, even at room temperature and typical densities. Thus, these properties make it a suitable medium to study the strong coupling state of matter. The medium properties in strongly coupled state-like collective structures \cite{tsytovich2004non, jana1993collective}, phase transitions \cite{maity2022parametric} have been observed. Dynamical equilibrium and chaotic dynamics of dust clusters have been illustrated recently \cite{deshwal2022chaotic,maity2020dynamical}. Thus, even homogeneous complex plasma has sought much community attention in a couple of decades.

    Generally, in industrial devices \cite{merlino2004dusty}  and natural environments \cite{merlino2021dusty}, dust particles are found to be polydisperse, which may be either of different shapes and sizes or composed of different types of materials \cite{block2019dusty}. Among the polydisperse systems, binary complex plasma is the simplest, consisting of only two types of particles. In recent years, there has been rapid development in the investigation of various phenomena in binary complex plasmas. Numerical simulations and experiments have been conducted to understand fluid separation \cite{ivlev2009fluid}, mixing \cite{jiang2011initial}, and demixing \cite{jiang2011demixing} in binary complex plasmas. A series of experiments on binary complex plasma has been done in the presence of micro-gravity using Plasma Kristall-4 (PK-4) \cite{thoma2006parabolic} experimental setup on board the International Space Station \cite{ivlev2009fluid}. There, phenomena like lane formation \cite{sutterlin2009lane} have been observed where smaller particles are injected into a cloud of bigger dust particles. Experiments on parabolic flights observed the phase separation in binary complex plasma systems with small size disparities of about $5 \%$ \cite{killer2016phase}. In ground-based laboratories, a 2D complex binary dust layer has been generated experimentally using two different types of materials \cite{wieben2017generation,matthews2006dynamics,assoud2008binary}. However, such a plasma is also susceptible to instabilities due to gradients in particle densities or in nonuniform applied fields. 
    
    In this article, we explore one of the instabilities in the binary complex plasma regime in the presence of an external electric field and equilibrium charge density gradient. Specifically, this work attempts to study electric field-driven Rayleigh-Taylor (RT)-like instability in dusty binary plasma by Langevin dynamics simulations using the large-scale atomic/molecular massively parallel simulator (LAMMPS) \cite{plimpton1995fast}. All particles are assumed to interact via a screened Coulomb potential $U(r) = (Q/4\pi\epsilon_0r)\exp{(-r/\lambda_D)}$ \cite{konopka2000measurement} and are confined by an external electric field. The charge-to-mass ratio of the two types of particles is kept the same so that all particles can levitate on a single plane inside the sheath. We will also try to understand the Rayleigh-Taylor instability analytically by deriving the expression for the growth rate in both weak and strong coupling regimes and shall validate this based on molecular dynamics simulations.     

	This paper is organized into five sections: In Section II, we analytically calculate the dispersion relation and growth rate for the RT-like instability in strong and weak coupling regimes. The details of Molecular Dynamics (MD) simulations for simulating binary complex plasma are provided in Section III. In Section IV, the matching of simulation and analytical results is shown. Finally, the conclusions of our work are given in Section V.

 \label{intro}


\section{Analytical results for electric field-driven RT-like instability in complex plasma}
\subsection{\it In weak coupling regime}\label{appA}
Considering plasma boundary in the y-z plane. We derive an expression for the growth rate of instability in the simplest case, making assumptions that (a) the effect of electrons and ions are taken in screened Coulomb potential; (b) the plasma is infinite in extent in $\hat{x}$ direction; (c) there is a density gradient in $\hat{y}$ direction; and (d) externally applied electric field is in $-\hat{y}$ direction. To solve this, we employ a linearization approach so that terms containing higher powers of amplitude can be neglected.  Additionally, we denote variables at equilibrium and those perturbed from equilibrium using subscripts $0$ and $1$, respectively, as written below:

\begin{equation}
\begin{split}
  \boldsymbol{v_d}= \boldsymbol{v_0} + \boldsymbol{v_1}; \quad 
  \boldsymbol{E}= {E_{0}(y)} \boldsymbol{\hat{y}}  + \boldsymbol{E_{1}(x,y)}; \quad 
  n = n_0+n_1 ; 
  \end{split}
\end{equation}

 Here, $\boldsymbol{v_d}$, $n$, and $\boldsymbol{E}$ represent the velocity, density of dust species and electric field, respectively. $\boldsymbol{E_0}$(y) is an externally applied electric field. 
 The momentum equation for the dust particles of mass $m_d$ is 
 \begin{equation}
m_d n \left(\frac{\partial \boldsymbol{v_d}} {\partial t} + (\boldsymbol{v_d}.\boldsymbol{\nabla})\boldsymbol{v_d} \right) = -Z_dn e \boldsymbol{E}  - \boldsymbol{\nabla} p  
\end{equation}

The equation can be rewritten in terms of charge $\rho_c=-Z_de$, $\alpha=k_BT_d/m_d$ and mass density $\rho_{md}=m_d n$
\begin{equation}
\rho_{m_{d}} \left(\frac{\partial \boldsymbol{v_d}} {\partial t} + (\boldsymbol{v_d}.\boldsymbol{\nabla})\boldsymbol{v_d} \right) = \rho_c \boldsymbol{E}  - \alpha \boldsymbol{\nabla} \rho_m  
\end{equation}

Putting $\rho_m= \beta \rho_c$  and $\bar{\alpha}= \alpha \beta$ where $\beta = -m_d/Z_d e$ we obtain 
\begin{equation}
\rho_{m_{d}} \left(\frac{\partial \boldsymbol{v_d}} {\partial t} + (\boldsymbol{v_d}.\boldsymbol{\nabla})\boldsymbol{v_d} \right) = \rho_c \boldsymbol{E}  - \bar{\alpha} \boldsymbol{\nabla} \rho_c  
\end{equation}
 In equilibrium, dust obeys the momentum equation 
  
\begin{equation}
\rho_{c0}{E_0} = \bar{\alpha} \boldsymbol{\nabla}\rho_{c0}
\end{equation}

The continuity can be written as
\begin{equation}
\frac{\partial \rho_{c}}{\partial t} + \boldsymbol{\nabla}. (\rho_{c} v_d)=0
\end{equation}
After perturbation, the linearized continuity and momentum equations for dust species are
\begin{equation}
\frac{\partial \rho_{c1}}{\partial t} + \boldsymbol{\nabla}. (\rho_{c0} v_1)=0
\end{equation}

\begin{equation}
\rho_{m0} \frac{\partial v_1}{\partial t} = \rho_1 \boldsymbol{E_0} + \rho_0 \boldsymbol{E_1} - \alpha \nabla \rho_{m1}  
\end{equation}

Electrons and ions follow the Boltzmann distribution 
\begin{equation}
n_{e} = n_{e0} exp\left({\frac{e\phi}{k_B T_e}}\right)
\end{equation}

\begin{equation}
n_{i} = n_{i0}  exp\left({\frac{-e\phi}{k_B T_i}}\right)
\end{equation}
The linearized Poisson's equation is
\begin{equation}
    \nabla^2 \phi_1 = \frac{1}{\lambda_D^2}\phi_1 + 4\pi \rho_{c1}
\end{equation} 
where $\lambda_D$ is Debye length given by 

$\lambda_D = \lambda_{De} \lambda_{Di}\sqrt{\lambda_{De}^2+\lambda_{Di}^2}$

   After linearizing the equation (7) we get the following equations: 
\begin{equation}
     \boldsymbol{k.v} = \frac{\omega \rho_{c1}}{\rho_{c0}} + \frac{\iota v_{1y} {\rho_{c0}}^{'}}{\rho_{c0}}
\end{equation}   
   
Linearizing the momentum equation $(8)$ along the $x$ and $y$ direction, we get,
\begin{equation}
    v_{1y} =  \frac{\rho_{c1} E_0}{\iota \rho_{m0} \omega } - \frac{k_y \rho_{c0} \phi_1}{\rho_{m0} \omega} -  \frac{\bar{\alpha} k_y \rho_{c1} }{\omega \rho_{m0}} 
\end{equation}
\begin{equation}
    v_{1x} = -\frac{k_x \rho_{c0} \phi_1}{n_0 m_d \omega} - \frac{\bar{\alpha} k_x \rho_{c1} }{\omega \rho_{m0}} 
\end{equation}
and Poisson's equation provides
\begin{equation}
    \phi_1 = \frac{4\pi \rho_{c1}}{k^2+k_D^2}
\end{equation}
where $k_D = \frac{1}{\lambda_D^2}$.
Now, multiply equations (13) and (14) by $k_y$ and $k_x$, respectively, and then add them to find
\begin{equation}
    \boldsymbol{k.v} = \frac{\rho_{c1} k_y E_0}{\iota \rho_{m0}\omega } - \frac{k^2 \rho_{c0} \phi_1}{\rho_{m0} \omega} - \frac{\bar{\alpha} k^2 \rho_{c1} }{\omega \rho_{m0}}
\end{equation}

Equating equations (12) and (16) by replacing $\phi_1$ and $v_{1y}$ using equations (15) and (13), respectively. We get the following dispersion relation
\begin{equation}
    \omega^2=\frac{E_0 (\rho_{c0}^{'}+\iota \rho_{c0} k_y)}{\rho_{m0}} +\left(1-\frac{\iota \rho_{c0}^{'}k_y}{\rho_{c0} k^2}\right) k^2 c_s^2
\end{equation}
where $c_s = \sqrt{\frac{1}{\rho_{m0}} \left(\frac{\rho_{c0}^2}{k^2 + k_D^2}+ \bar{\alpha} \rho_{c0}\right)}$ is dust acoustic speed.

\begin{figure}[hbt!]
    \includegraphics[height = 10.0cm,width = 9.0cm]{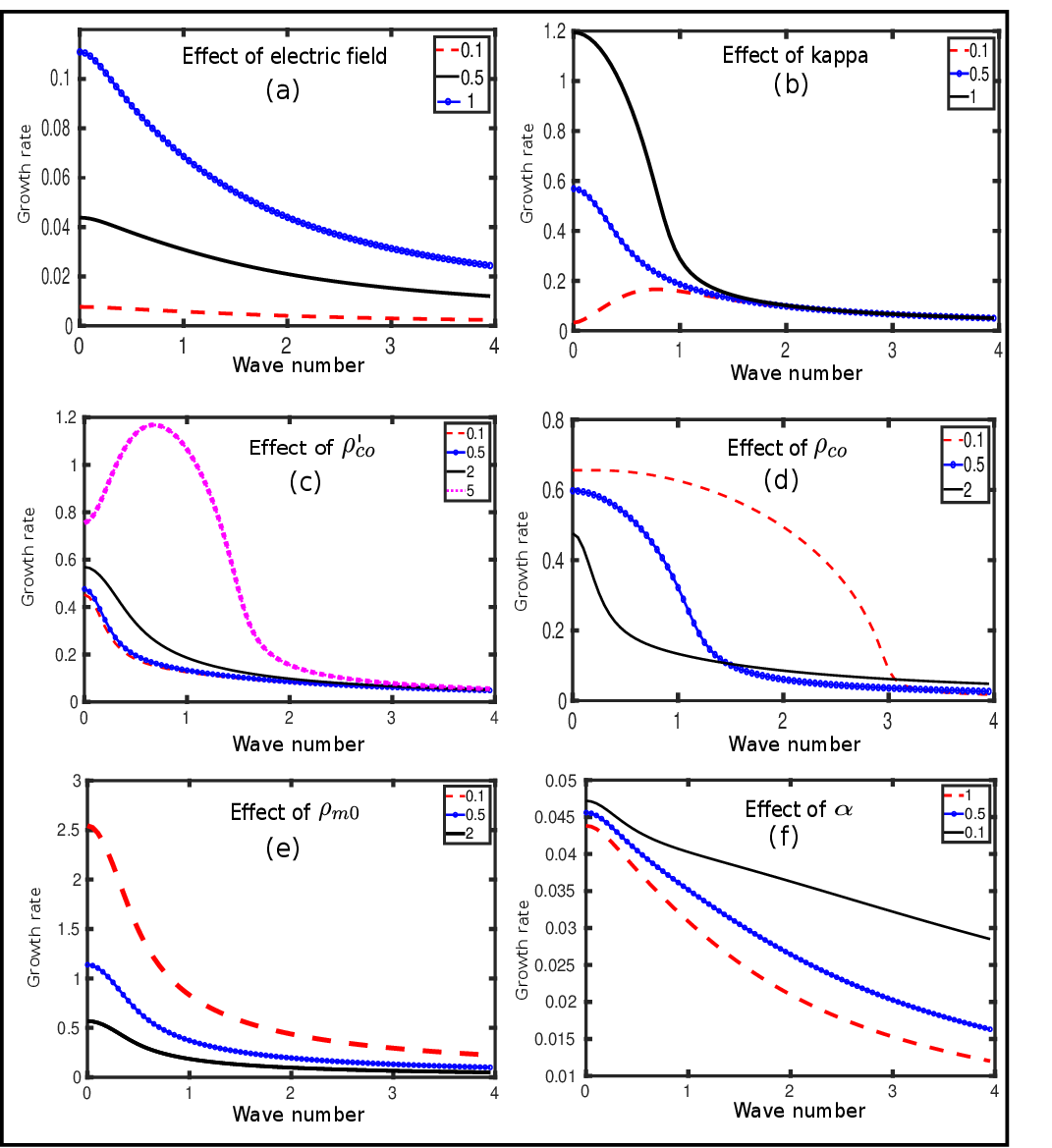}
    \caption{Variation of growth rate with wave number on changing various parameters. Each subplot examines the impact of one parameter while holding all other variables constant. Different colors within a single subplot visually represent the influence of adjusting a particular parameter.}
\label{figl1}
\end{figure}

In Fig.\ref{figl1}, we have employed various subplots to illustrate how the growth rate changes with wave number ($k_y$) by modifying distinct parameters. Each subplot shows the effects of a single parameter while keeping all other parameters constant, and different colors in a single subplot depict the influence of adjusting a specific parameter whose effect we have to observe.

The growth rate changes when different parameters are altered while keeping others constant. From the subplot (a) of Fig.\ref{figl1}, which shows the effect of the electric field, it is clear that as the electric field increases, the growth rate increases. The externally applied electric field provides a driving force to instability, so as the electric field increases, the driving electric force increases, and as a result, the growth rate of instability increases. Subplot (b) illustrates the effect of kappa, where we note that the growth rate is greater for the higher values of kappa for a given value of perturbation ($k_y$). This is because at very high kappa ($\kappa_D$) values, the pair interaction between the particles becomes weak, and this weakening of inter-particle correlations makes the medium less resistant to perturbations, allowing instabilities to grow more easily. When the initial charge density ($\rho_{c0}$) increases, the growth rate of the instability increases, which is shown in subplot (d) of Fig.\ref{figl1} because an increase in the initial dust charge density enhances the dust plasma frequency and reduces Debye shielding, thus strengthening field-particle coupling and promoting instability growth, especially in the presence of equilibrium drifts or gradients. Similarly, we obtain the variation of the growth rate with parameters like $\rho_{m0}$, $\alpha$, and ${\rho^{'}_{c0}}$. An increase in the mass density of particles suppresses the growth rate of instability because increasing mass increases the inertia of particles, which reduces the response of particles to perturbation. 
    
    In the next section, we will consider the strong coupling effect and derive the dispersion relation using the Generalized Hydrodynamic model (GHD) \cite{diaw2015generalized, kaw1998low}.

\subsection{\it In strong coupling regime using the generalized hydrodynamics model}

Dusty plasma can be considered as a viscoelastic fluid in a coupling parameter $1>\Gamma>175$ regime. Viscoelastic fluids have traits of both solids and liquids, and the Generalized Hydrodynamics Model (GHD) \cite{kaw1998low} is a phenomenological model to explain the dynamics of such fluids. 
According to the GHD model \cite{diaw2015generalized, rosenberg1997dust}, the momentum equation of negatively charged dust species can be written as
\begin{equation}  
\begin{split}
\left(1+\tau_m \frac{\partial}{\partial t} \right)\left(m_d n_d\frac{\partial\boldsymbol{v_d}}{\partial t} + {\boldsymbol{\nabla}P} + Z_d e n_d \boldsymbol{E}\right) = \\ \eta\nabla^2\boldsymbol{ v_d} + \left(\frac{\eta}{3}+\xi \right) \boldsymbol{\nabla(\nabla.v_d)}
\end{split}
\end{equation}

where quantities $n_d$ and $v_d$ are the density and velocity of the dust fluid, respectively. The equation of state gives pressure $P$=$\mu_d n_d k_B T_d$ with $T_d$ and $\mu_d$ as the temperature and the compressibility coefficient of dust species. Here $\tau_m$ is the memory relaxation time, and $\eta$ and $\zeta$ are, respectively, the shear and bulk viscosities of the dusty plasma medium.

We can rewrite the above equation as
\begin{equation}  
\begin{split}
\left(1+\tau_m \frac{\partial}{\partial t} \right) \left(\rho_{md}\frac{\partial\boldsymbol{v_d}}{\partial t} + {\bar{\alpha}\boldsymbol{\nabla}\rho_c} - \rho_c \boldsymbol{E}\right) = \\ \eta\nabla^2\boldsymbol{ v_d} + \left(\frac{\eta}{3}+\xi \right) \boldsymbol{\nabla(\nabla.v_d)}
\end{split}
\end{equation}

The linearized equation of motion is 
 \begin{equation}  
\begin{split}
\left(1+\tau_m \frac{\partial}{\partial t} \right) \left(\rho_{m0}  \frac{\partial\boldsymbol{v_1}}{\partial t} + {\bar{\alpha}\boldsymbol{\nabla}\rho_{c1}}  -\rho_{c0} \boldsymbol{E_1} - \rho_{c1} \boldsymbol{E_0}\right) = \\ \eta^* \nabla^2\boldsymbol{ v_1} 
\end{split}
\end{equation} 
Here, $ \eta ^{*} = \left( \frac{4\eta}{3} + \xi \right)$.  
  In equilibrium, the momentum equation becomes 
\begin{equation}
\rho_{c0}{E_0} = \bar{\alpha} \boldsymbol{\nabla}\rho_{c0}
\end{equation}

After perturbation, the linearized form of continuity and momentum equations for dust species are
\begin{equation}
\frac{\partial \rho_{c1}}{\partial t} + \nabla. (\rho_{c0} v_1)=0
\end{equation}

\begin{equation}
    \nabla^2 \phi_1 = \frac{1}{\lambda_D^2}\phi_1 + 4\pi \rho_{c1}
\end{equation} 

The x and y components of the momentum equation are
\begin{equation}  
\begin{split}
v_{1x} = \frac{-\iota k_x\frac{\rho_{c1}}{\rho_{co}}\left( \frac{\rho_{c0}^2}{k^2+k_D^2} + \bar{\alpha}\rho_{c0} \right)}{\left( -\iota \rho_{m0} + \frac{\eta^* k^2}{(1-\iota \omega \tau_m)}\right)}
\end{split}
\end{equation} 

\begin{equation}  
\begin{split}
 v_{1y} =  \frac{\rho_{c1}}{\rho_{c0}}\frac{ \left[\rho_{c0} E-\iota k_y\left( \frac{\rho_{c0}^2}{k^2+k_D^2} + \bar{\alpha}\rho_{c0} \right)\right]}  {\left( -\iota \rho_{m0} + \frac{\eta^* k^2}{(1-\iota \omega \tau_m)}\right)}
\end{split} 
\end{equation}

From the Fourier transform of the continuity equation, we get
\begin{equation}  
\begin{split}
-\iota \omega \rho_1 + \rho_0^{'} v_{1y} + \iota \rho_0 \left( \boldsymbol{k.v}\right) = 0
\end{split}
\end{equation}
Now, using equations (24-26), we obtain the following dispersion relation
\begin{equation}  
\begin{split}
1= \frac{-\iota k^2\left( \frac{\rho_{c0}^2}{k^2+k_D^2}+ \bar{\alpha}\rho_{c0}\right)-\rho_{c0}E_0 k_y}{\omega\left(-\iota \omega \rho_{m0}+ \frac{\eta^* k^2}{(1-\iota \omega \tau_{m})}\right)} + \\ \frac{\iota \rho_{c0}^{'}}{\rho_{c0} \omega}\frac{-\iota k_y\left( \frac{\rho_{c0}^2}{k^2+k_D^2}+ \bar{\alpha}\rho_{c0}\right)}{\left(-\iota \omega \rho_{m0}+ \frac{\eta^* k^2}{(1-\iota \omega \tau_{m})}\right)}
\end{split}
\end{equation}
For a weak coupling regime, $\tau_m$ becomes zero, and viscous forces become much smaller than inertial forces, and we obtain the dispersion relation the same as equation (17).
\begin{figure}[hbt!]
   \includegraphics[height = 4.5cm,width = 6.0cm]{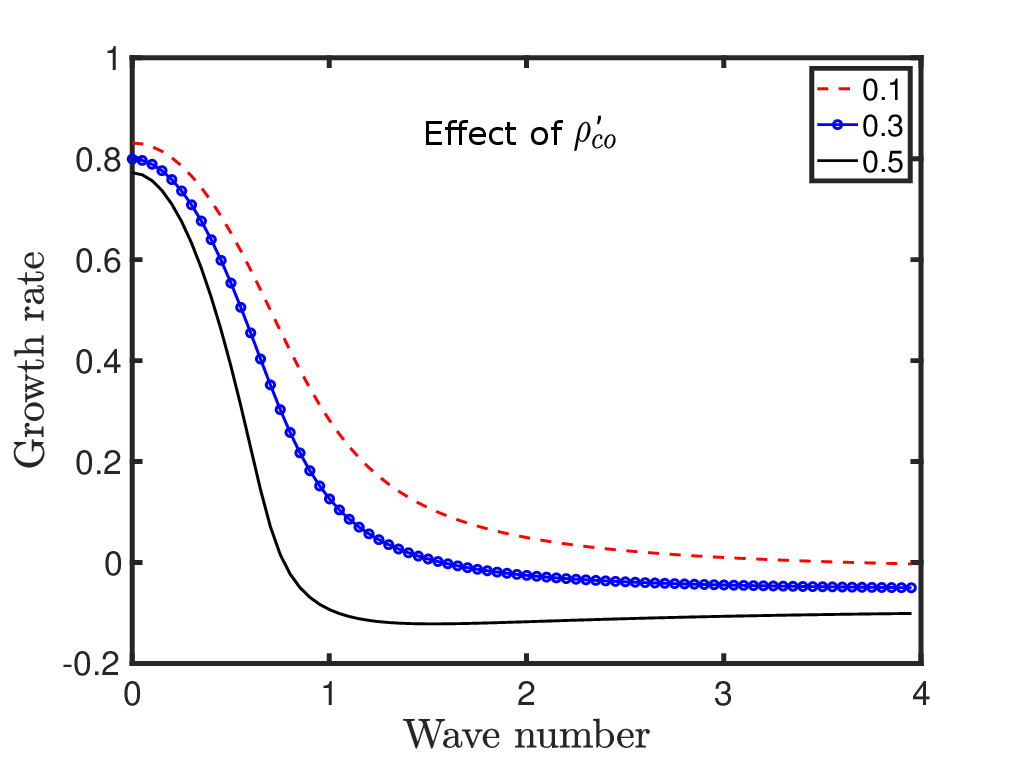}
   \caption{{Variation of growth rate with wave number on changing charge density while keeping all parameters constant ($\rho_{c0}$).}  }
  \label{figa3l}
\end{figure}

\begin{figure}[hbt!]
   \includegraphics[height = 10.0cm,width = 9.0cm]{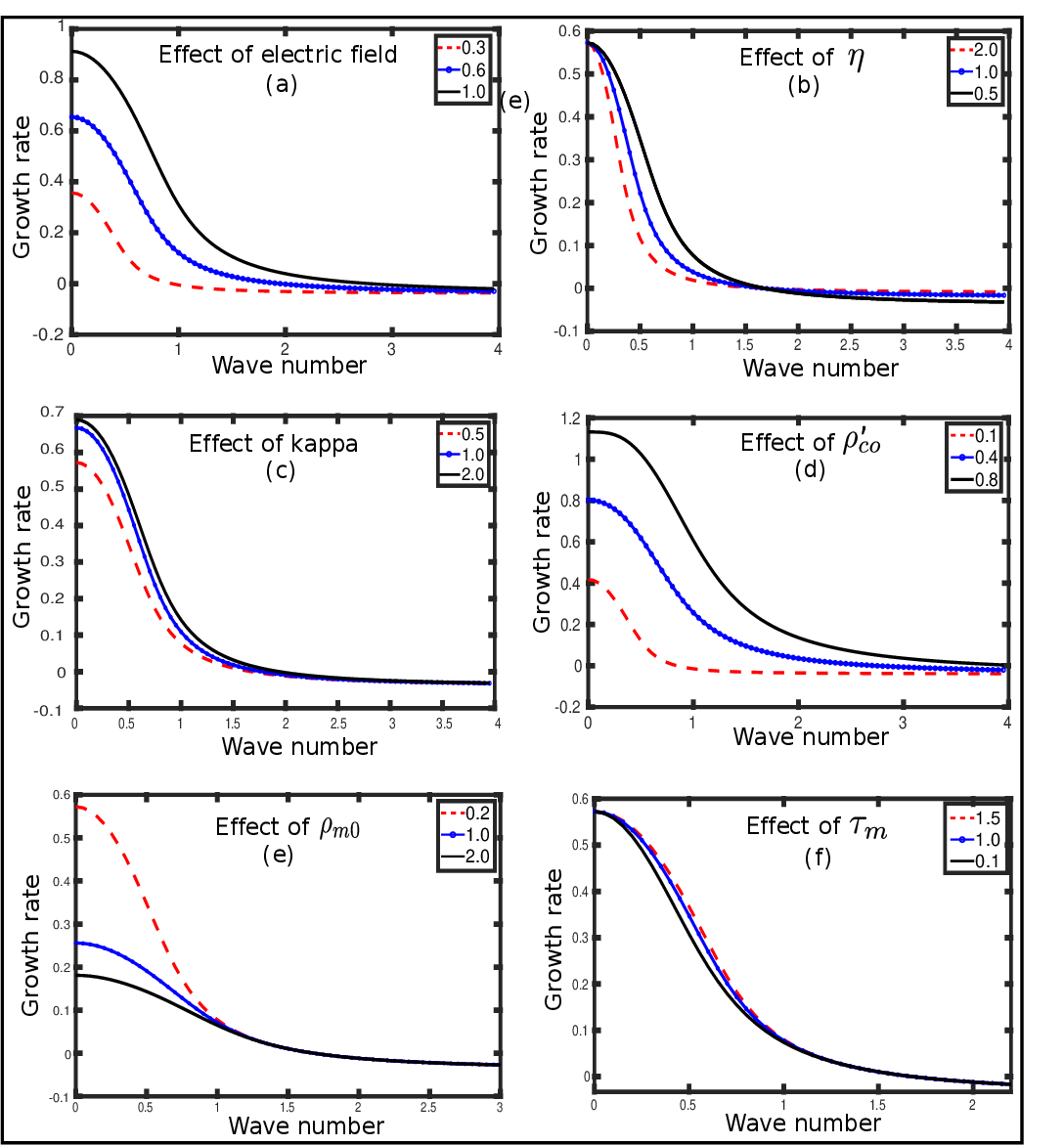}
   \caption{{Variation of growth rate with wave number on changing various parameters. Each subplot examines the impact of one parameter while holding all other variables constant. Different colors within a single subplot visually represent the influence of adjusting a particular parameter.}  }
  \label{figa2l}
\end{figure}

According to Fig.\ref{figa3l}, the growth rate decreases as the charge density ($\rho_{c0}$) increases;  this is because increasing the charge density leads to a stronger inter-particle Coulomb interaction. This enhances the Coulomb coupling parameter ($\Gamma$), pushing the system further into the strongly coupled regime. As a result, the medium exhibits an increased viscoelastic behavior, characterized by stronger elastic restoring forces and long-range correlations. These effects act to resist perturbations, thereby suppressing the growth of instability. The variation of the growth rate with wavenumber for different parameter values is shown in Fig.\ref{figa2l}. Parameters electric field, kappa, $\rho_{m0}$, and ${\rho^{'}_{c0}}$ follow the same trend as explained in the weakly coupled regime.
It can be seen from subplot (f) of Fig.\ref{figa2l} that on increasing the relaxation time ($\tau_m$), the growth rate of instability increases. $\tau_m$
 characterizes the memory of the viscoelastic strongly coupled medium. A higher value of $\tau_m$ leads to slower relaxation of stress and a more elastic response of the dust component. Consequently, the energy associated with perturbations is stored rather than dissipated, which enhances the instability growth rate. On increasing viscosity coefficient ($\eta$), energy dissipation increase, which in turn reduce perturbation energy, resulting in a decrease of growth rate, which is shown in subplot (b) of Fig.\ref{figa2l}.

    In the next section, molecular dynamics simulations are conducted to validate theoretical predictions of Rayleigh-Taylor (RT) instability in a binary dusty plasma system. 

\section{\it MD Simulation Details}

In this study, 2D molecular dynamics simulations of binary complex plasma crystals are conducted using an open-source MD code, LAMMPS \cite{plimpton1995fast}, in which all the charged particles interact via a screened Coulomb potential and are confined by an external electric field, $E_0$ in calculations. Mass $(M_{1})$ and charge $(Q_{1})$ of one of the species of dust particles that are created above and below the central region are taken as $6.99\times 10^{-13}$kg and $11940e$ \cite{nosenko2004shear}. Particles having mass $M_{2} = 0.1M_{1}$ and charge $Q_{2}=0.1Q_{1}$ are created in the central region of the simulation box with constrain that the charge-to-mass ratio of both the species remains the same. Each particle in the simulation box interacts via screened Coulomb potential, $U(r) = (Q/4\pi\epsilon_0r)\exp{(-r/\lambda_D)}$. $\lambda_D$ representing the characteristic screening length for the Yukawa interaction between particles. Length scales are normalized by the debye length $\lambda_0 = 2.2854 \times 10^{-3} \, \text{m}$. Accordingly, the dimensionless screening parameter, which characterizes the strength of the pair interaction, is defined as
\begin{equation}
    \kappa = \frac{\lambda_0}{\lambda_D}.
\end{equation}
Dust particles are confined by an external force, which is provided by an externally applied electric field of the form $\mathbf{E}_y =  K(y-L/2) \hat{y}$. Here, $K$ is the strength of the confining potential. Dynamics of dust particles are tracked by choosing a time step $dt = 0.0001 \omega_{pd}^{-1}$ for simulations where $\omega_{pd}^{-1}=(nQ_f^2/\epsilon M_f)^{-1/2} = 0.29079 s^{-1}$.

 A 2D simulation box is created having lengths $L= L_x = L_y = 12.7943\lambda_D$ in the x- and y-directions, respectively, and periodic boundary conditions in \text{x}-direction.

 Langevin's equation of motion for each particle is 
\begin{equation}
 m_i\ddot{\mathbf{r}_{i}} = -Q\sum_{j=1}^{N_p} \nabla U(\mathbf{r}_i, \mathbf{r}_j) - \nu m_i \dot{\mathbf{r_i}}  + Q \mathbf{E}_{y} + \xi_i(t)	
\end{equation} 	  
Here, the first term represents the force that arises due to the interaction of an $i^{th}$ a particle with others, and $\nu m_i \dot{r_i}$ term includes the frictional drag of neutrals, where $\nu$ is the friction coefficient, and $\xi_i (t)$ is a random force having a Gaussian distribution with zero mean, i.e., $ <\xi_i (t)>$ = 0, and variance $ <\xi_i(t) \xi_i (t+\tau)> = 2 \nu_i m_i k_B T \delta(\tau)$. Here, $ \delta $ stands for the Dirac Delta function and $k_B$ is the thermal energy of the bath. This random force mimics the random kicks of background neutrals acting on each particle. We have continued our simulations for quite a long time.

\label{mdsim}

\section{\it  Results and discussion}

\begin{figure}[hbt!]
   \includegraphics[height = 5.5cm,width = 6.0cm]{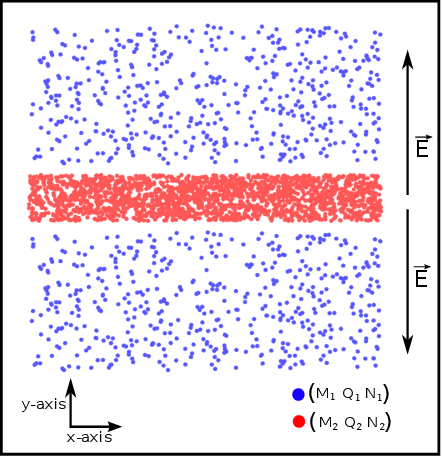}
   \caption{Initial configuration of particles in the simulation box.  }
  \label{figal4}
\end{figure}

\begin{figure}[hbt!]
   \includegraphics[height = 6.0cm,width = 7.0cm]{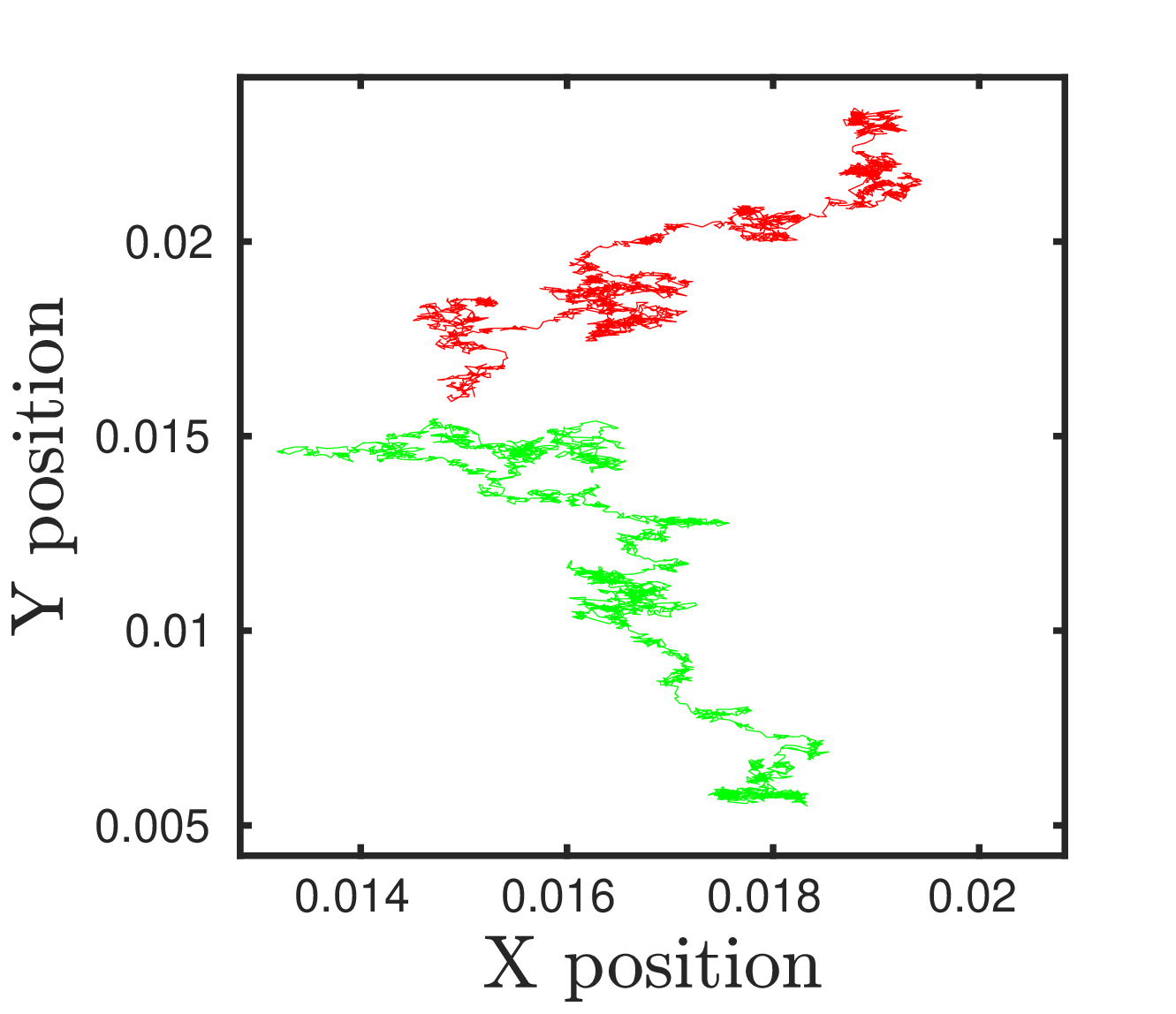}
   \caption{Time evolution of the trajectory of two particles in the simulation box.  }
  \label{figal5}
\end{figure}

We created $N=3000$ particles in the simulation box with periodic boundary conditions in the $x$-direction. The charged particles $N_1$=2000 created in the rectangular region near the center of the simulation box, and surrounding particles are created $ N_2$=1000 above and below the central region, shown by blue and red dots, respectively, in Fig.\ref{figal4}. We fix the charge ($Q_1$) and mass ($M_1$) of the surrounding species, and the central region species have charge and mass $ 0.1Q_1 $ and $ 0.1M_1$, respectively. Now two distinct regions get formed with charge densities $9.7\times10^{-7}$ $Cm^{-2}$ of the central region and $7.7\times10^{-7}$ $Cm^{-2}$ of regions above and below the central region, respectively. As mentioned earlier, the particles are subjected to an external electric field of the form $\mathbf{E_y} =  K(y-L/2) \hat{y}$. We choose the damping parameter $\nu=10$ and strength of pair interaction $\kappa=1$. Both species feel different forces as an externally applied force depends on the charge and position of the particles.  The less charged species drift towards the outer edge to gain a minimum energy configuration. In Fig.\ref{figal5}, we have shown the trajectory of two particles of the central regions drifting towards the edges at $\Gamma=300 $. Here, red and green represent the trajectories of two central region particles drifting upward and downward, respectively. Clearly, particles in the central region drift toward the edges of the simulation box due to the external electric field. When the heavier dust species is positioned above the lighter one, the configuration becomes dynamically unstable due to the effective acceleration induced by the electric field. This instability promotes the amplification of interfacial perturbations, ultimately leading to the development of RT-like structures and interspecies mixing. 

\begin{figure}[hbt!]
   \includegraphics[height = 6.0cm,width = 9.0cm]{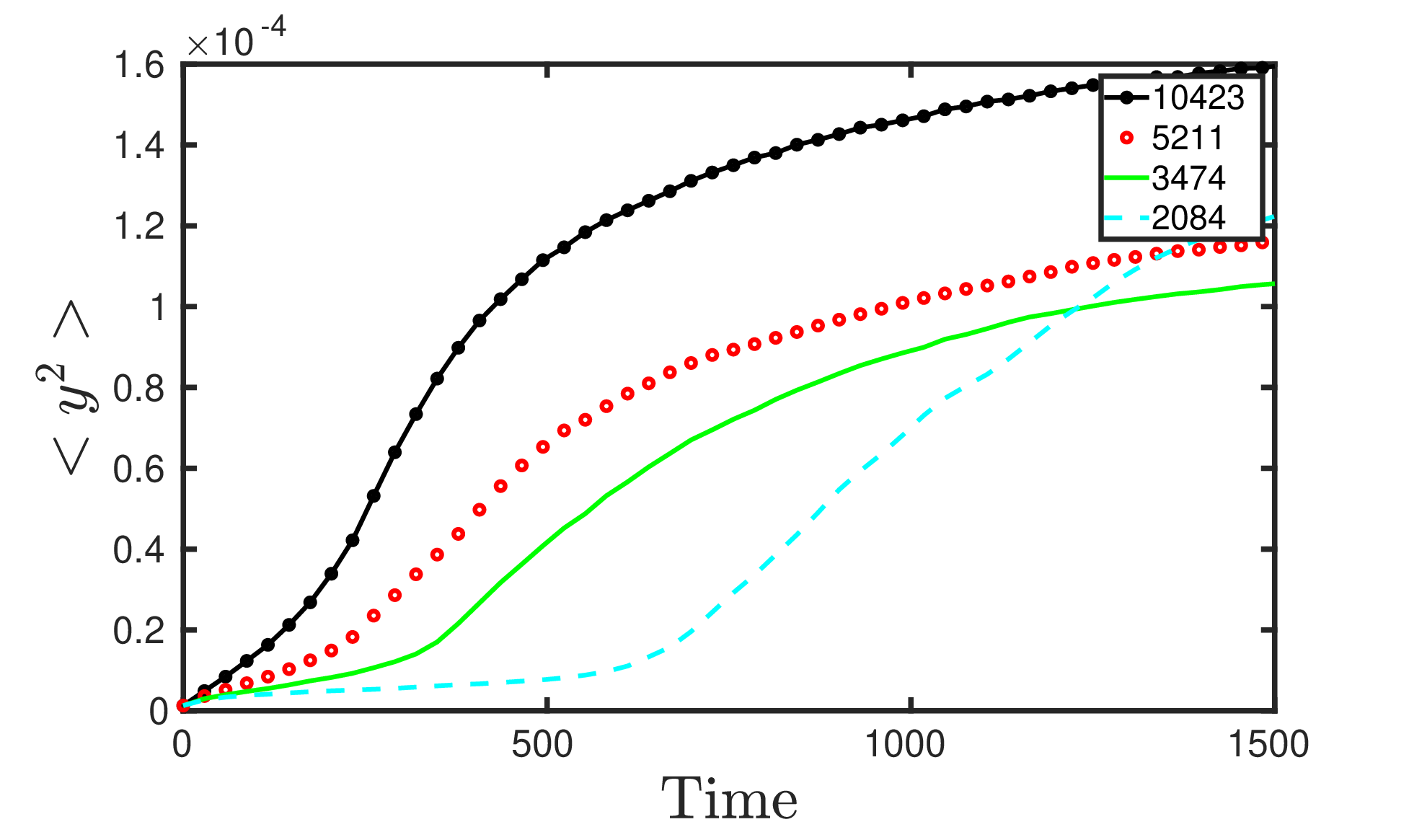}
   \caption{Time evolution of mean square displacement ($<y^2>$) of central region particles for different values of temperature.}
  \label{figal6}
\end{figure}

The temporal evolution of the mean square displacement in the vertical direction, denoted as $\langle y^2 \rangle$, for particles initially located in the central region (depicted in red), is presented in Fig.~\ref{figal6} across a range of temperatures. In this figure, the curves corresponding to the highest and lowest temperatures are represented by green and cyan colors, respectively. At higher temperatures, the initial behavior of $\langle y^2 \rangle$ exhibits an approximately linear trend, indicative of rapid vertical displacement driven by an externally applied electric field and initial charge density. As the temperature decreases, the slope of this linear regime diminishes, reflecting a reduction in particle mobility. This trend can be attributed to the increase in the Coulomb coupling parameter $\Gamma$ with decreasing temperature, which leads to stronger interparticle correlations and a more strongly coupled system.

Furthermore, each $\langle y^2 \rangle$ curve eventually reaches a saturation point over longer timescales. This saturation arises as lighter particles, initially confined to the central region, migrate vertically and penetrate through the heavier particle layer due to the combined effects of electrostatic interactions and thermal motion. Over time, the system reaches a quasi-steady state where most of the lighter particles have moved toward the peripheries of the simulation domain. A fraction of these particles, however, become immobilized or jammed in interstitial regions formed by surrounding heavier particles, leading to incomplete relaxation. This dynamical behavior highlights the complex interplay between thermal energy, interspecies mass asymmetry, and Coulomb interactions in determining the vertical transport and mixing characteristics within a dusty plasma environment.

\begin{figure}[hbt!]
   \includegraphics[height = 6.0cm,width = 9.0cm]{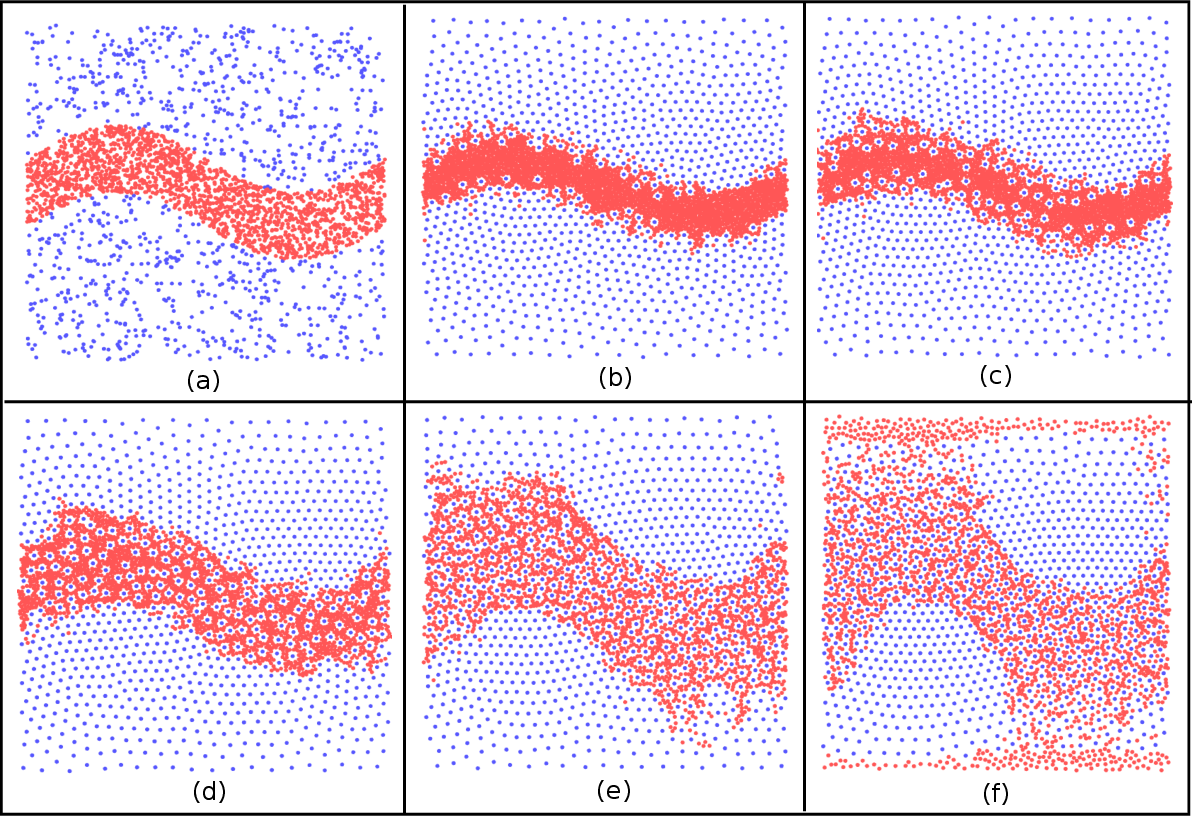}
   \caption{Time evolution of a sine-shaped central region. Here, subplot (a) represents the initial position of particles inside the simulation box, and the others show snapshots at different times during evolution.}
  \label{figa7l}
\end{figure}

To understand the problem more efficiently, we have simulated the evolution of a sine-shaped central region consisting of particles having charge $Q_1$ and mass $M_1$, respectively. Concurrently, we also consider other particles with charges and masses labeled as $Q_2$ and $M_2$ in the neighboring vicinity of this central region. Fig.\ref{figa7l} shows the time evolution of the sine-shaped region. Subplot (a) of Fig.\ref{figa7l} shows the initial position of particles inside the simulation box, and the subplots from (a) to (f) show how the sine-shaped region evolves with time.
\begin{figure}[hbt!]
   \includegraphics[height = 7.5cm,width = 8.0cm]{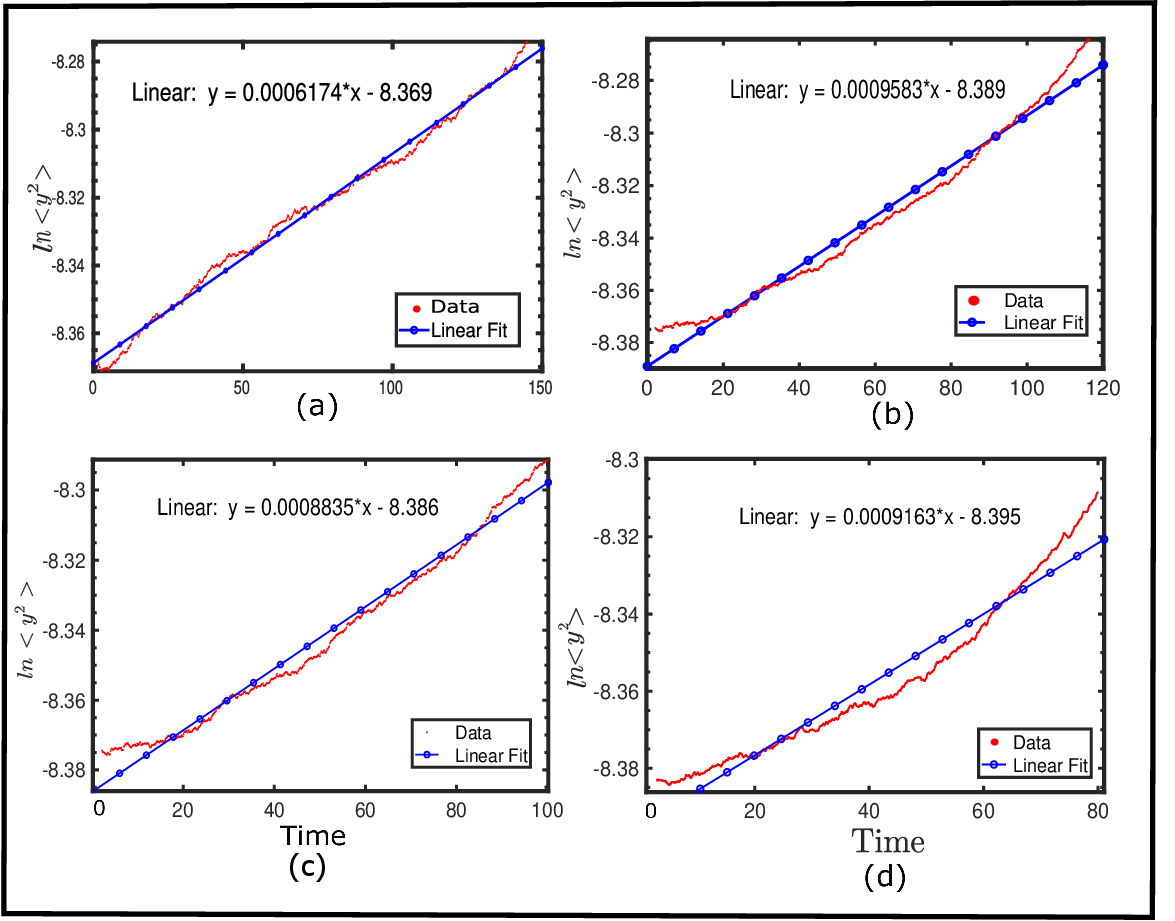}
   \caption{Time evolution of $\ln \langle y^2 \rangle$ for particles in the central region. Subplots (a) $K = 2500$, $\kappa = 1$; (b) $K = 3000$, $\kappa = 1$ illustrates the effects of strength of electric field parameter $K$; and (c) $\kappa = 2$, $K = 2500$(d) $\kappa = 2.5$,  $K = 2500$ shows the effect of particle interactions. Here, the red color plots depicts the simulation data, and the blue color line represents the linear fitted curve.}
  \label{figa8l}
\end{figure}

The growth rate ($\gamma$) of instability is extracted from the time evolution of the mean square interface displacement. In the linear regime, $y^2(t) =y_o^2 exp(2\gamma t)$ exhibits exponential growth, allowing it to be obtained from the slope of ln $y^2$(t) versus time. Only early times are considered, for which the interface amplitude remains small compared to the perturbation wavelength and nonlinear effects are negligible. In Fig.\ref{figa8l}, we have shown the time evolution of ln$(y^2)$ and the slope of the curve, i.e. $2\gamma$, which gives a growth rate of instability. In the simulations, the wavelength of the perturbed sine wave is $2\pi/0.03 = 210 $, where $0.03$ is the length of the simulation box. In subplot (a) of Fig.\ref{figa8l}, the red color plot depicts the simulation data, and the blue color line represents the linear fitted curve. The slope of the fitted (blue) curve is 0.0006174, which corresponds to the value of $\gamma =  0.0003087$. 

\begin{figure}[hbt!]
   \includegraphics[height = 6.0cm,width = 8.0cm]{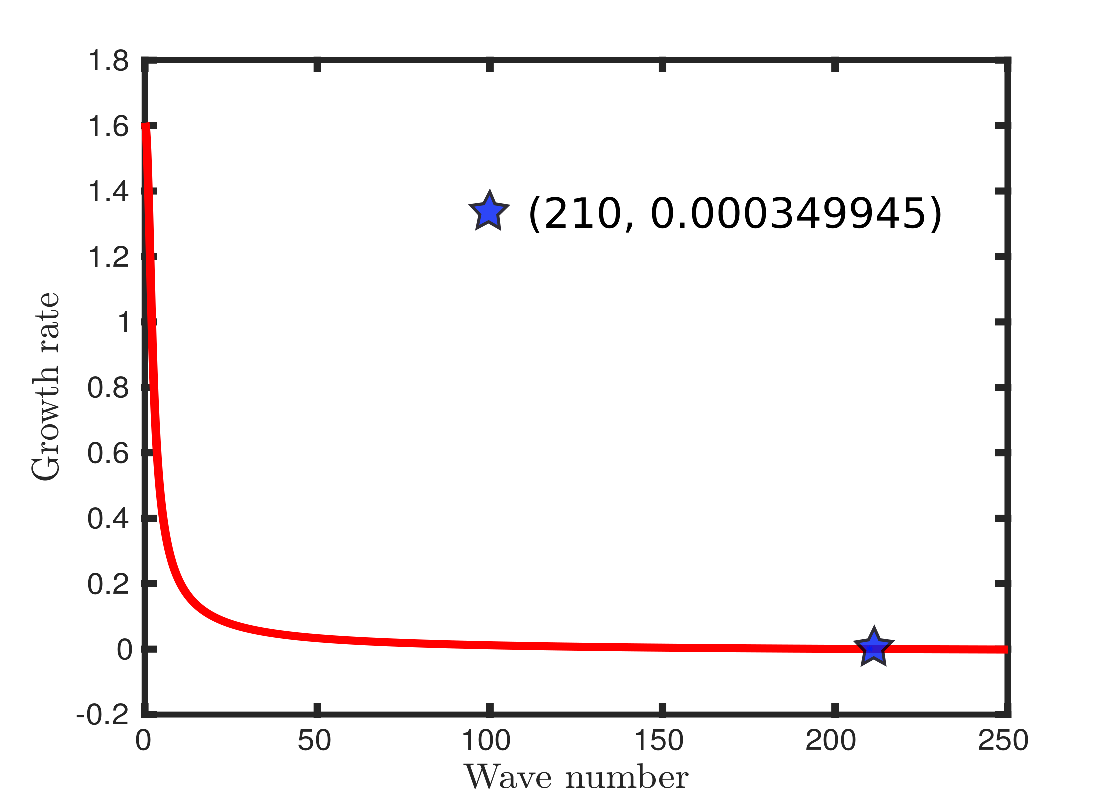}
   \caption{Growth rate as a function of wavenumber for a specific set of parameters.}
  \label{figa9l}
\end{figure}

Using the dispersion relation (27), we have plotted the growth rate (imaginary part of $\omega$) as a function of wave number for a given set of parameters in Fig.\ref{figa9l}. The growth rate of the perturbation corresponding to the wavenumber \( k = 210 \) is calculated to be \( 0.000349945 \), which is in excellent agreement with the slope extracted from the simulation data presented in Fig.~\ref{figa8l}. This consistency between the theoretical prediction and the numerical results validates the analytical model and confirms the reliability of the simulation approach.

We have also performed simulations to investigate the effect of varying the strength of the confining potential by altering the value of $K$, as illustrated in Subplots (a) and (b) of Fig.~\ref{figa8l}. The results indicate that increasing the confining potential strength leads to a higher growth rate of instability, which is consistent with the analytical findings presented in Fig.~\ref{figa2l}. Additionally, simulations conducted for different values of the screening parameter ($\kappa$), shown in Subplots (c) and (d) of Fig.~\ref{figa8l}, demonstrate that an increase in $\kappa$ also results in an enhanced growth rate. These outcomes are in agreement with the analytical results in Fig.~\ref{figa2l}. Similar trends are observed for other parameters, such as particle charge and mass density, which also show good agreement with analytical predictions.

\section{\it  Summary}
This research work analyzed Rayleigh–Taylor–like instability in a strongly coupled binary complex plasma system using two-dimensional Langevin dynamics simulations and modeled interfacial disturbances between two species of charged dust particles having the same charge-to-mass ratio under an externally applied electric field. The findings indicate that lesser/lighter charged particles migrate across regions of higher/haveier charge, driven by the electric field, resulting in the emergence of instability. Analytical approaches have been used to determine the growth rate of instability in both weakly and strongly coupled plasma regimes, with these results validated by molecular dynamics simulations. We present an entire parametric analysis, enhancing the understanding of RT instability dynamics in complex plasmas while providing a basis for subsequent experimental and theoretical studies.

 \begin{center}
    \textbf{ACKNOWLEDGEMENTS} 
 \end{center}
  The authors thank the IIT Delhi HPC facility for computational resources. Priya Deshwal is thankful to the University Grants Commission [Grant No. 09/086(1413)/2019-EMR-I] for funding the research.
\bibliography{cluster_ref}

@article{konopka2000measurement,
  title={Measurement of the interaction potential of microspheres in the sheath of a rf discharge},
  author={Konopka, U and Morfill, GE and Ratke, L},
  journal={Physical review letters},
  volume={84},
  number={5},
  pages={891},
  year={2000},
  publisher={APS}
}

@article{lampe2000interactions,
  title={Interactions between dust grains in a dusty plasma},
  author={Lampe, Martin and Joyce, Glenn and Ganguli, Gurudas and Gavrishchaka, Valeriy},
  journal={Physics of Plasmas},
  volume={7},
  number={10},
  pages={3851--3861},
  year={2000},
  publisher={American Institute of Physics}
}

@article{barkan1996experiments, 
title={Experiments on ion-acoustic waves in dusty plasmas},
  author={Barkan, A and D'angelo, N and Merlino, RL},
  journal={Planetary and Space Science},
  volume={44},
  number={3},
  pages={239--242},
  year={1996},
  publisher={Elsevier}
}

@article{rao1990dust,
title={Dust-acoustic waves in dusty plasmas},
  author={Rao, NN and Shukla, PK and Yu, M Yu},
  journal={Planetary and space science},
  volume={38},
  number={4},
  pages={543--546},
  year={1990},
  publisher={Elsevier}
}

@article{shukla1992dust,
  title={Dust ion-acoustic wave},
  author={Shukla, PK and Silin, VP},
  journal={Physica Scripta},
  volume={45},
  number={5},
  pages={508},
  year={1992}
}

@article{ma1994self,
  title={Self-consistent theory of ion acoustic waves in a dusty plasma},
  author={Ma, Jin-Xiu and Yu, MY},
  journal={Physics of Plasmas},
  volume={1},
  number={11},
  pages={3520--3522},
  year={1994},
  publisher={American Institute of Physics}
}

@article{thomas1994plasma,
  title={Plasma crystal: Coulomb crystallization in a dusty plasma},
  author={Thomas, H and Morfill, GE and Demmel, V and Goree, J and Feuerbacher, B and M{\"o}hlmann, D},
  journal={Physical Review Letters},
  volume={73},
  number={5},
  pages={652},
  year={1994},
  publisher={APS}
}

@article{rosenberg1996ion,
  title={Ion-dust streaming instability in processing plasmas},
  author={Rosenberg, M},
  journal={Journal of Vacuum Science \& Technology A: Vacuum, Surfaces, and Films},
  volume={14},
  number={2},
  pages={631--633},
  year={1996},
  publisher={American Vacuum Society}
}

@article{melzer1994experimental,
  title={Experimental determination of the charge on dust particles forming Coulomb lattices},
  author={Melzer, Andr{\'e} and Trottenberg, Thomas and Piel, Alexander},
  journal={Physics Letters A},
  volume={191},
  number={3-4},
  pages={301--308},
  year={1994},
  publisher={Elsevier}
}

@article{pieper1996experimental,
  title={Experimental studies of two-dimensional and three-dimensional structure in a crystallized dusty plasma},
  author={Pieper, JB and Goree, J and Quinn, RA},
  journal={Journal of Vacuum Science \& Technology A: Vacuum, Surfaces, and Films},
  volume={14},
  number={2},
  pages={519--524},
  year={1996},
  publisher={American Vacuum Society}
}

@article{tsytovich2004non,
  title={Non-linear collective phenomena in dusty plasmas},
  author={Tsytovich, VN and Morfill, GE},
  journal={Plasma physics and controlled fusion},
  volume={46},
  number={12B},
  pages={B527},
  year={2004},
  publisher={IOP Publishing}
}

@article{jana1993collective,
  title={Collective effects due to charge-fluctuation dynamics in a dusty plasma},
  author={Jana, MR and Sen, A and Kaw, PK},
  journal={physical review E},
  volume={48},
  number={5},
  pages={3930},
  year={1993},
  publisher={APS}
}

@article{merlino1998laboratory,
  title={Laboratory studies of waves and instabilities in dusty plasmas},
  author={Merlino, RL and Barkan, A and Thompson, C and D’angelo, N},
  journal={Physics of Plasmas},
  volume={5},
  number={5},
  pages={1607--1614},
  year={1998},
  publisher={American Institute of Physics}
}

@article{d1993rayleigh,
  title={The Rayleigh-Taylor instability in dusty plasmas},
  author={d'Angelo, N},
  journal={Planetary and space science},
  volume={41},
  number={6},
  pages={469--474},
  year={1993},
  publisher={Elsevier}
}

@article{d1990kelvin,
  title={The Kelvin-Helmholtz instability in dusty plasmas},
  author={d'Angelo, N and Song, Bin},
  journal={Planetary and space science},
  volume={38},
  number={12},
  pages={1577--1579},
  year={1990},
  publisher={Elsevier}
}

@article{kaw1998low,
  title={Low frequency modes in strongly coupled dusty plasmas},
  author={Kaw, PK and Sen, A},
  journal={Physics of Plasmas},
  volume={5},
  number={10},
  pages={3552--3559},
  year={1998},
  publisher={American Institute of Physics}
}

@article{kaw2001collective,
  title={Collective modes in a strongly coupled dusty plasma},
  author={Kaw, PK},
  journal={Physics of Plasmas},
  volume={8},
  number={5},
  pages={1870--1878},
  year={2001},
  publisher={American Institute of Physics}
}

@article{ivlev2000anisotropic,
  title={Anisotropic dust lattice modes},
  author={Ivlev, AV and Morfill, G},
  journal={Physical Review E},
  volume={63},
  number={1},
  pages={016409},
  year={2000},
  publisher={APS}
}

@article{maity2022parametric,
  title={Parametric decay induced first-order phase transition in two-dimensional Yukawa crystals},
  author={Maity, Srimanta and Arora, Garima},
  journal={Scientific Reports},
  volume={12},
  number={1},
  pages={1--11},
  year={2022},
  publisher={Nature Publishing Group}
}

@article{deshwal2022chaotic,
  title={Chaotic dynamics of small-sized charged Yukawa dust clusters},
  author={Deshwal, Priya and Yadav, Mamta and Prasad, Chaitanya and Sridev, Shantam and Ahuja, Yash and Maity, Srimanta and Das, Amita},
  journal={Chaos: An Interdisciplinary Journal of Nonlinear Science},
  volume={32},
  number={6},
  pages={063136},
  year={2022},
  publisher={AIP Publishing LLC}
}

@article{maity2020dynamical,
  title={Dynamical states in two-dimensional charged dust particle clusters in plasma medium},
  author={Maity, Srimanta and Deshwal, Priya and Yadav, Mamta and Das, Amita},
  journal={Physical Review E},
  volume={102},
  number={2},
  pages={023213},
  year={2020},
  publisher={APS}
}

@article{pieper1996dispersion,
  title={Dispersion of plasma dust acoustic waves in the strong-coupling regime},
  author={Pieper, JB and Goree, J},
  journal={Physical review letters},
  volume={77},
  number={15},
  pages={3137},
  year={1996},
  publisher={APS}
}

@article{morfill1996plasma,
  title={Plasma crystal},
  author={Morfill, GE and Thomas, H},
  journal={Journal of Vacuum Science \& Technology A: Vacuum, Surfaces, and Films},
  volume={14},
  number={2},
  pages={490--495},
  year={1996},
  publisher={American Vacuum Society}
}

@article{block2019dusty,
  title={Dusty (complex) plasmas—routes towards magnetized and polydisperse systems},
  author={Block, Dietmar and Melzer, Andre},
  journal={Journal of Physics B: Atomic, Molecular and Optical Physics},
  volume={52},
  number={6},
  pages={063001},
  year={2019},
  publisher={IOP Publishing}
}

@article{diaw2015generalized,
  title={Generalized hydrodynamics model for strongly coupled plasmas},
  author={Diaw, Abdourahmane and Murillo, Michael Sean},
  journal={Physical Review E},
  volume={92},
  number={1},
  pages={013107},
  year={2015},
  publisher={APS}
}

@article{rosenberg1997dust,
  title={Dust acoustic waves in strongly coupled dusty plasmas},
  author={Rosenberg, M and Kalman, G},
  journal={Physical Review E},
  volume={56},
  number={6},
  pages={7166},
  year={1997},
  publisher={APS}
}

@article{ivlev2009fluid,
  title="\MYhref{https://iopscience.iop.org/article/10.1209/0295-5075/85/45001/meta?casa_token=kpDO_sqhxZUAAAAA:Fb35pEUeMN1irhCAJmAlvqKp35QiJCbAGL9ZZYKS12nEadqD9ZHi6_MbDO6InLnuZoCieRHWbTChqpl9_DAfIEad7oHo}{Fluid phase separation in binary complex plasmas}",
  author={Ivlev, AV and Zhdanov, SK and Thomas, HM and Morfill, GE},
  journal={Europhysics Letters},
  volume={85},
  number={4},
  pages={45001},
  year={2009},
  publisher={IOP Publishing}
}

@article{jiang2011initial,
  title={Initial stages in phase separation of binary complex plasmas: Numerical experiments},
  author={Jiang, K and Hou, L-J and Ivlev, AV and Li, Y-F and Du, C-R and Thomas, HM and Morfill, GE and S{\"u}tterlin, KR},
  journal={Europhysics Letters},
  volume={93},
  number={5},
  pages={55001},
  year={2011},
  publisher={IOP Publishing}
}

@article{jiang2011demixing,
  title={Demixing in binary complex plasma: computer simulation},
  author={Jiang, Ke and Hou, Lu-Jing and Ivlev, Alexei V and Li, Yang-Fang and Sutterlin, K Robert and Thomas, Hubertus M and Morfill, Gregor E},
  journal={IEEE Transactions on Plasma Science},
  volume={39},
  number={11},
  pages={2752--2753},
  year={2011},
  publisher={IEEE}
}

@article{thoma2006parabolic,
  title={Parabolic flight experiments with PK-4},
  author={Thoma, MH and H{\"o}fner, H and Kretschmer, M and Ratynskaia, S and Morfill, GE and Usachev, A and Zobnin, A and Petrov, O and Fortov, V},
  journal={Microgravity-Science and Technology},
  volume={18},
  pages={47--50},
  year={2006},
  publisher={Springer}
}

@article{merlino2004dusty,
  title={Dusty plasmas in the laboratory, industry, and space},
  author={Merlino, Robert L and Goree, John A},
  journal={PHYSICS TODAY.},
  volume={57},
  number={7},
  pages={32--39},
  year={2004},
  publisher={AMERICAN INSTITUTE OF PHYSICS}
}

@article{merlino2021dusty,
  title={Dusty plasmas: from Saturn’s rings to semiconductor processing devices},
  author={Merlino, Robert},
  journal={Advances in Physics: X},
  volume={6},
  number={1},
  pages={1873859},
  year={2021},
  publisher={Taylor \& Francis}
}

@article{sutterlin2009lane,
  title={Lane formation in driven binary complex plasmas on the international space station},
  author={Sutterlin, K Robert and Thomas, Hubertus M and Ivlev, Alexei V and Morfill, Gregor E and Fortov, Vladimir E and Lipaev, Andrey M and Molotkov, Vladimir I and Petrov, Oleg F and Wysocki, Adam and Lowen, Hartmut},
  journal={IEEE Transactions on Plasma Science},
  volume={38},
  number={4},
  pages={861--868},
  year={2009},
  publisher={IEEE}
}

@article{killer2016phase,
  title={Phase separation of binary charged particle systems with small size disparities using a dusty plasma},
  author={Killer, Carsten and Bockwoldt, Tim and Sch{\"u}tt, Stefan and Himpel, Michael and Melzer, Andr{\'e} and Piel, Alexander},
  journal={Physical review letters},
  volume={116},
  number={11},
  pages={115002},
  year={2016},
  publisher={APS}
}

@article{wieben2017generation,
  title={Generation of two-dimensional binary mixtures in complex plasmas},
  author={Wieben, Frank and Schablinski, Jan and Block, Dietmar},
  journal={Physics of Plasmas},
  volume={24},
  number={3},
  pages={033707},
  year={2017},
  publisher={AIP Publishing LLC}
}

@article{plimpton1995fast,
  title={Fast parallel algorithms for short-range molecular dynamics},
  author={Plimpton, Steve},
  journal={Journal of computational physics},
  volume={117},
  number={1},
  pages={1--19},
  year={1995},
  publisher={Elsevier}
}

@article{nosenko2004shear,
  title={Shear flows and shear viscosity in a two-dimensional Yukawa system (dusty plasma)},
  author={Nosenko, V and Goree, J},
  journal={Physical review letters},
  volume={93},
  number={15},
  pages={155004},
  year={2004},
  publisher={APS}
}

@article{matthews2006dynamics,
  title={Dynamics of a dust crystal with two different size dust species},
  author={Matthews, Lorin Swint and Qiao, Ke and Hyde, Truell Wayne},
  journal={Advances in Space Research},
  volume={38},
  number={11},
  pages={2564--2570},
  year={2006},
  publisher={Elsevier}
}

@article{assoud2008binary,
  title={Binary crystals in two-dimensional two-component Yukawa mixtures},
  author={Assoud, Lahcen and Messina, Ren{\'e} and L{\"o}wen, Hartmut},
  journal={The Journal of chemical physics},
  volume={129},
  number={16},
  year={2008},
  publisher={AIP Publishing}
}
\end{document}